\documentclass[11pt,a4paper]{article}
\usepackage{jheppub}
\usepackage{amsmath}
\usepackage{graphicx}
\usepackage{color}
\begin{document}

\title{Charmonium spectral functions from 2+1 flavour lattice QCD}

\author[1]{Szabolcs Bors\'anyi,}
\author[1,2]{Stephan D\"urr,}
\author[1,2,3]{Zolt\'an Fodor,}
\author[1]{Christian Hoelbling,}
\author[3,4]{S\'andor D. Katz,}
\author[1,2]{Stefan Krieg,}
\author[5]{Simon Mages,}
\author[3,4]{D\'aniel N\'ogr\'adi,}
\author[3,4]{Attila P\'asztor,}
\author[5]{Andreas Sch\"afer,}
\author[1,2]{K\'alm\'an K. Szab\'o,}
\author[1]{B\'alint C. T\'oth,}
\author[3,4]{Norbert Trombit\'as}

\affiliation[1]{University of Wuppertal, Department of Physics, Wuppertal D-42097, Germany}
\affiliation[2]{J\"ulich Supercomputing Center, J\"ulich D-52425, Germany}
\affiliation[3]{E\"otv\"os University, Budapest 1117, Hungary}
\affiliation[4]{MTA-ELTE Lend\"ulet Lattice Gauge Theory Research Group}
\affiliation[5]{University of Regensburg, Regensburg D-93053, Germany}

\abstract{
Finite temperature charmonium spectral functions in the pseudoscalar and vector 
channels are studied in lattice QCD with 2+1 flavours of dynamical Wilson quarks, on fine 
isotropic lattices (with a lattice spacing of $0.057\rm{fm}$), 
with a non-physical pion mass of $m_{\pi} \approx 545 \rm{MeV}$. 
The highest temperature studied is
approximately 1.4$T_c$. Up to this temperature no significant variation of the
spectral function is seen in the pseudoscalar channel. The vector channel shows
some temperature dependence, which seems to be consistent with a temperature dependent
low frequency peak 
related to heavy quark transport, plus a 
temperature independent term at $\omega>0$. These results are in accord with previous 
calculations using the quenched approximation.
}

\maketitle

\section{Introduction}

\subsection{Motivation}
As the early universe expanded after the Big Bang, a transition occurred at $T \approx 150$ MeV, 
from the so called quark-gluon plasma to a confined phase with hadrons. The nature of this 
transition affects our understanding of the history of the universe, see e.g. \cite{Schwarz:2003du}. \\

Extensive experimental work is currently done at both RHIC and LHC, trying to understand the QCD
aspects of the early universe with heavy ion collisions. Both for the cosmological transition and
for these heavy ion experiments, the densities are quite small, and so the baryonic chemical
pontentials are much less than the typical hadron masses. Therefore a $\mu=0$ calculation is
a good approximation for both cases. \\

The $\mu=0$ transition can be studied using lattice gauge theory. There are various results
using different fermion regularizations. Within the staggered formalism there are continuum extrapolated
results at physical quark masses, such as the nature of the transition, the transition temperature, equation of state and fluctuations
\cite{Aoki:2006we,Aoki:2005vt,Aoki:2009sc,Borsanyi:2010cj,Bazavov:2011nk,Borsanyi:2011sw,Bazavov:2012jq}. There are also results with the 
computationally more expensive, but conceptually cleaner Wilson fermions
\cite{Umeda:2008bd, Ejiri:2009hq, Maezawa:2009nd, Bornyakov:2009qh, Umeda:2010ye, Maezawa:2011aa, Umeda:2012er}, 
but so far only at non-physical masses, here however, continuum results are also available \cite{WilsonThermo}. In this paper, 
we use Wilson fermions, since spectroscopy is easier to handle \cite{Science},
and we will need to measure correlators at finite temperatures. We use a fine lattice spacing of $a=0.057$ fm,
but continuum extrapolation does not seem feasible at this point.\\

The charmonium systems have been under heavy investigation, since the suppression of $J/\Psi$
is regarded as an important experimental signal for the formation of the plasma 
state \cite{CharmoniumOriginalPhenom}. Charmonium states are expected to dissolve somewhat
above the transition temperature, because of the screening of the inter-quark potential and
collisions with the hot medium. In this paper, we investigate the in medium properties of 
the $J/\Psi$ and $\eta_c$ mesons from a lattice QCD perspective. \\

\subsection{Mesonic spectral functions}

The spectral function (SF) of a correlator of self-adjoint operators is the imaginary part of the Fourier-transform of the 
real time retarded correlator \cite{LeBellac}. 

\begin{equation}
A_H(\omega)=\frac{1}{\pi} \Im D^R_H(\omega, \vec{p} ) = \frac{1}{2 \pi} \left( D^{>}_H(\omega, \vec{p}) - D^{<}_H(\omega, \vec{p} ) \right) \rm{,}
\end{equation}
\begin{equation}
D^{>(<)}_H(t,\vec{x}) = \int \frac{d^4p}{(2 \pi)^4} e^{ipx} D^{>(<)}_H(\omega,\vec{p}) \rm{,} \;\;\; \rm{where}
\end{equation}
\begin{equation}
D^{>}_H(t,\vec{x}) = \left< J_H(t,\vec{x}) J_H(0,\vec{0})\right> \ \rm{,} \ t>0 \;\;\; \rm{ and}
\end{equation}
\begin{equation}
D^{<}_H(t,\vec{x}) = \left< J_H(0,\vec{0}) J_H(t,\vec{x})\right> \ \rm{,} \ t>0 \rm{.}
\end{equation}

In this article, we will deal with correlators between mesonic currents, and 
the corresponding mesonic SFs. These operators schematically look 
like 
\begin{equation}
J_H(t,\vec{x}) = \bar{q}(\vec{x},t) \Gamma_H q(\vec{x},t)\rm{,} 
\end{equation}
where $q$ is the quark field
and $\Gamma_H = \gamma_5, \gamma_i$ for the pseudoscalar and vector channels
respectively. \\

It can be shown, that the SF is related to the Euclidean correlator --
calculable on the lattice -- by an integral transform
\begin{equation}
\label{eq:IntegralTransform}
G(\tau,\vec{p}) = \int_0^\infty d\omega A(\omega,\vec{p}) K(\omega,\tau) \rm{,}
\end{equation}
where we dropped the subscript H, 
\begin{equation}
K(\omega,\tau) = \frac{\cosh(\omega(\tau-1/{2T}))}{\sinh(\omega/{2T})}
\end{equation}
is the integral kernel, and the Euclidean correlator (at zero chemical potential) is:
\begin{equation}
G(\tau,\vec{p}) = D^{>}_H(-i\tau,\vec{p}) = \int d^3x e^{i\vec{p}\vec{x}} \left< T_{\tau} J_H(-i\tau,\vec{x}) J_H(0,\vec{0}) \right> \rm{.}
\end{equation}

Knowledge of these SFs is of great importance. Inserting a
complete set of states and using the so-called Kubo--Martin--Schwinger condition,
one can show that:
\begin{equation}
A(\omega,\vec{p}) = \frac{(2\pi)^2}{Z} \sum_{m,n} \left( e^{-E_n/T} - e^{-E_m/T}\right) \left| \left< n | J_H(0) | m\right> \right|^2 \delta^{(4)}(p_{\mu} - k^n_{\mu} + k^m_{\mu}) \rm{,}
\end{equation}
where $k_i=(E_i,\vec{k}_i)$ and $p=(\omega,\vec{p})$. 
In this sum, a stable particle gives a $\delta$ like peak \cite{LeBellac}, while an unstable particle in matter
gives a smeared peak. Also, an important result of linear response theory  is the Kubo-formula,
which states that the transport coefficients are related to the zero frequency limit of $A(\omega)/\omega$.
In particular, the heavy quark diffusion constant D is related to the vector spectral function:
\begin{equation}
\label{equ:Kubo}
D=\frac{1}{6 \chi} \lim_{\omega \to 0} \sum_{i=1}^3 \frac{\rho_{ii}(\omega,T)}{\omega}  \rm{,}
\end{equation}
where $\chi$ is the (heavy) quark number susceptibility and $\rho_{ii}$ is the spectral function corresponding to the vector channel. 
If the transport coefficient is non vanishing, we expect some finite value of $\rho/\omega$ for small $\omega$. This implies the presence of a transport peak.
We will investigate the anticipated melting of the heavy meson states $J/\Psi$ and $\eta_c$ in the
quark gluon plasma, which is supposed to happen somewhat above the transition temperature \cite{CharmoniumOriginalPhenom}.
As one increases the temperature, the width of a given peak increases and at sufficiently 
high temperatures, the contribution from the meson state in the SF may 
be sufficiently broad so that it is not very meaningful to speak of it as a well defined 
state any more. If the peak corresponding to the given states disappears from 
the SF in this way, we can say it melted. \\

In the following, we will only be dealing with SFs at zero spatial momentum $A(\omega,\vec{0})=A(\omega)$, but we
point out that the analysis of the non-zero spatial momentum SFs would go the same way, except one would need to start
from non-zero spatial momentum correlators.\\

\subsection{The Maximum Entropy Method}
To get the SFs from a lattice study one has to invert equation (\ref{eq:IntegralTransform}). This 
inversion however is ill-defined, since the typical number of frequencies for which one
wants to reconstruct the SF is higher 
than the number of data points. In this case a $\chi^2$ fit on the shape of the SF discretized
to $N_\omega$ points is degenerate. One has to regularize the problem in some way. 
\footnote{Note that the number
of data points vs. parameters is not the only problem when trying to reconstruct SFs. Even if one has a large number
of data points (e.g. on time anisotropic lattices), the Euclidean correlator is rather insensitive to fine details of
the SF. Therefore the inversion introduces large uncertainties \cite{Aarts:2002cc}.}\\

The determination of hadronic SFs via the Maximum Entropy Method (MEM) was 
first suggested in \cite{Asakawa}. This is a regularization that has some justification from
Bayesian probability theory. One has to maximize:
\begin{equation}
\label{eq:Q}
Q = \alpha S - \frac{1}{2} \chi^2  \rm{,}
\end{equation}
where as usual
\begin{equation}
\chi^2 = \sum_{i,j=1}^{N_{\rm{data}}} (G_i^{\rm{fit}}-G_i^{\rm{data}}) C^{-1}_{ij} (G_j^{\rm{fit}}-G_j^{\rm{data}}) \rm{,}
\end{equation}
with $C_{ij}$ being the covariance matrix of the data (in Euclidean time), and the Shannon--Jaynes entropy\footnote{This is the generalization of the 
Shannon entropy to continuous probability distributions, it is the negative of the Kulback-Leibler divergence.} is
\begin{equation}
S = \int d\omega \left( A(\omega)-m(\omega)-A(\omega) \log\left(\frac{A(\omega)}{m(\omega)}\right) \right) \rm{,}
\end{equation}
where the so called prior function $m(\omega)$ is supposed to summarize our prior knowledge on the shape of the 
SF (such as the leading perturbation theory behaviour). It can then be shown that the maximum of $Q$ lies in an $N_{\rm{data}}$ dimensional subspace of
the $N_\omega$ dimensional space of possible $A(\omega_i)$ vectors, that can be parametrized as:
\begin{equation}
\label{eq:subspace}
A(\omega)=m(\omega) \exp\left( \sum_{i=1}^{N_{\rm{data}}} s_i f_i(\omega)\right) \rm{.}
\end{equation}

The most widely used choice for basis functions involves a Singular Value Decomposition and is 
called the Bryan method. It was introduced in \cite{Bryant}. The particular choice of the 
basis for the subspace we use  is $f_i(\omega)=K(\omega,\tau_i)$ and was introduced 
by Ref. \cite{Jakovac}. In our experience this proved to be numerically more stable 
than the former one. In this case the maximization of Q is equivalent to the minimization of 
\begin{equation}
\label{eq:U}
U=\frac{\alpha}{2} \sum_{i,j=1}^{N_{\rm{data}}} s_i C_{ij} s_j + \int_0^{\omega_{\rm{max}}}\rm{d}\omega  A(\omega) - \sum_{i=1}^{N_{\rm{data}}} G_i^{\rm{data}} s_i \rm{.}
\end{equation}

One can see from equation (\ref{eq:subspace}) that the shape of the subspace is strongly dependent on the
choice of the prior function. This is the source of a systematic uncertainty, that has to be considered. \\

After equation (\ref{eq:U}) is minimized at a given value of $\alpha$, and the optimal $A_\alpha$ is obtained, 
the regularization parameter $\alpha$ has to be averaged over. Using Bayes' theorem, 
one can show that the conditional probability of $\alpha$ having a specific value, given the 
data and the prior function is \cite{Asakawa}:
\begin{equation}
P[\alpha|D,m] \propto \exp \left( \frac{1}{2}\sum_k \log \frac{\alpha}{\alpha+\lambda_k} + \alpha S - \frac{1}{2} \chi^2 \right) \rm{,}
\end{equation}
where the $\lambda$-s are the eigenvalues of the matrix 
$\Lambda_{l,{l'}} = \frac{1}{2} \left( \sqrt{A_l} \frac{\partial^2 \left( \chi^2 \right) }{\partial A_l \partial A_{l'}} \sqrt{A_{l'}} \right)_{A=A_\alpha} $.
The efficient marginalization of $\alpha$ involves a trick using Sylvester's determinant theorem, that can
be found in the appendix of \cite{Rothkopf}. \\

We also mention, that we use a modified version of the kernel and the spectral functions for the reconstruction \cite{Aarts:2007wj, Engels09}:
\begin{equation}
\hat{K}\left(\tau,\omega\right)= \tanh \left( \omega/2 \right) K \left( \tau, \omega \right) \rm{,}
\end{equation}
and
\begin{equation}
\hat{A} \left( \omega \right) = \coth \left( \omega/2 \right) A \left( \omega \right) \rm{.}
\end{equation}
This ``cures" the low frequency divergent $1/\omega$ behaviour of the kernel, without spoiling the high $\omega$
behaviour. In the rest of the paper, we will only be dealing with Euclidean time, and we will simply denote it with $t$ (changing the notation from $\tau$). \\ 

Lattice studies of charmonium SFs using the MEM have been carried out on numerous occasions
 (\cite{Jakovac, Umeda05, Asakawa04, Aarts07, Iida06, Ohno11, Ding, Aarts, Kelly:2013cpa}), but 
so far not in 2+1 flavour QCD. A recent, detailed study of charmonium SFs in 
quenched QCD can be found in \cite{Ding}. Results regarding spectral functions 
with 2 flavours of dynamical quarks can be found in Refs. \cite{Aarts, Kelly:2013cpa}. 
A recent study of electric conductivity using 2+1 flavours of anisotropic Wilson fermions
can be found in \cite{Amato:2013naa}. Another interesting application of the spectral function reconstruction 
is the study of the melting bottonium states in the context of Non-relativistic QCD \cite{Aarts:2013kaa}. 
Some details about the numerical implementation and the error analysis can be found in Appendix A.

\subsection{Lattice configurations}

We use the same lattice configurations as in \cite{WilsonThermo}.
The gauge action used for the calculations was the Symanzik tree level
improved gauge action \cite{Symanzik:1983dc, Luscher:1984xn}
\begin{equation}
S_G^\mathrm{Sym}  =  \beta\,\,\Big[\frac{c_0}{3}\,
\sum_\mathrm{plaq} {\rm Re\, Tr\,}(1-U_\mathrm{plaq})
\, + \frac{c_1}{3}\, \sum_\mathrm{rect} {\rm Re \, Tr\,}
(1- U_\mathrm{rect})\Big],
\end{equation}
with the parameters $c_0 = 5/3$ and $c_1= -1/12$. The action for the
fermionic sector was the clover improved \cite{Sheikholeslami:1985ij}
Wilson action
\begin{equation}
S_F^{\mathrm{SW}} = S_F^\mathrm{W}-
\frac{c_{\mathrm{SW}}}{4}\,\sum_x\,\sum_{\mu,\nu}
\bar{\psi}_x\,\sigma_{\mu\nu}F_{\mu\nu,x}\,\psi_x\,,
\end{equation}
where $S_F^\mathrm{W}$ is the Wilson fermion action. Six steps of stout
smearing \cite{Morningstar:2003gk} with smearing parameter $\varrho=0.11$ were
used. 
The clover coefficient was set to
its tree level value, $c_{\mathrm{SW}}=1.0$, which, for this type of smeared
fermions, essentially leads to an ${\cal O}(a)$ improved action
\cite{Hoffmann:2007nm} with improved chiral properties \cite{Capitani:2006ni}.
The same action was first used in Ref.~\cite{Durr:2008rw} where the excellent
scaling properties of hadron masses was observed. The full hadron spectrum
using this action was determined in Ref.~\cite{Science}.

The bare masses of the $u$ and $d$ quarks were taken to be degenerate,
therefore the configurations were generated using an $N_f=2+1$ flavor
algorithm. The $u$ and $d$ quarks were implemented via the Hybrid Monte Carlo (HMC)
algorithm~\cite{Duane:1987de}, whereas the strange quark was implemented using
the Rational Hybrid Monte Carlo (RHMC) algorithm~\cite{Clark:2006fx}. In order
to speed up the molecular dynamics calculations, the Sexton-Weingarten
multiple time-scale integration scheme \cite{Sexton:1992nu} combined with the
Omelyan integrator \cite{Takaishi:2005tz} was employed.  When all four extents
of the lattice were even, the usage of even-odd preconditioning
\cite{DeGrand:1988vx} gave an additional speed up factor of 2. 

\begin{table}[t!]
\centering 
\begin{tabular}{c c c c c c c c c}
\hline\hline 
$a[\rm{fm}]$ & $am_{ud}$ & $am_s$ & $ m_{\pi} $ & $N_s$ &  $N_t$ & $T = \frac{1}{N_t a}$  & \# confs. \\ 
\hline 
0.057(1) & -0.00336 &0.0050 & 545MeV & 64 & 28 &  123MeV & 151 \\
0.057(1) & -0.00336 &0.0050 & 545MeV & 64 & 20 &  173MeV & 95  \\
0.057(1) & -0.00336 &0.0050 & 545MeV & 64 & 18 &  192MeV & 328 \\
0.057(1) & -0.00336 &0.0050 & 545MeV & 64 & 16 &  216MeV & 254 \\
0.057(1) & -0.00336 &0.0050 & 545MeV & 64 & 14 &  247MeV & 411 \\
0.057(1) & -0.00336 &0.0050 & 545MeV & 64 & 12 &  288MeV & 300 \\
\hline
\hline
\end{tabular}

\caption{The different lattices used in the study. The configurations are separated by 5 trajectories.
We have explicitly checked, 
that none of the covariance matrices $C_{ij}$ of the data we have
display pathological spectra.}

\end{table}

From the study in \cite{WilsonThermo}, we only used the finest lattices, with
gauge coupling $\beta=3.85$, corresponding to a lattice spacing of $a=0.057(1)\rm{fm}$.
The bare light quark masses where choosen to be $am_{ud}=-0.00336$ and $am_s=0.0050$, which,
when fixing the scale with a physical $\Omega$ baryon mass, corresponds to a pion mass
$m_{\pi} \approx 545 \rm{MeV}$. A summary of the lattices used can be found in Table 1.

\begin{table}[t!]
\begin{center}
\begin{tabular}{l l l l l l l}
\hline\hline 
$J^P$ & $m_i$ & name & ma & $ma/m_{D_s^{*}}a$ & $m_{exp}[MeV]$ & $m_{exp}/m_{D_s^{*}}$ \\ 
\hline 
$0^-$ 		& $m_s$,$m_c$		& $D_s$		& 0.54(1)   & 0.95(2)	& 1968.4 	& 0.932 \\
$0^-$ 		& $m_c$,$m_c$ 		& $\eta_c$	& 0.8192(7) & 1.437(4)	& 2981.0 	& 1.411	\\
$1^-$ 		& $m_s$,$m_c$ 		& $D_s^*$	& 0.570(1)  & 1		& 2112.3	& 1 \\
$1^-$ 		& $m_c$,$m_c$ 		& $J/\Psi$	& 0.8388(8) & 1.472(2)	& 3096.916	& 1.466	\\
${3/2}^{+}$ 	& 3$m_s$ 		& $\Omega$ 	& 0.478(8)  & 0.84(2)	& 1672.45	& 0.791\\

\hline\hline
\end{tabular}
\end{center}

\caption{The different hadron masses, obtained by fitting $A \cosh(ma(t-N_t/2))$ to smeared 
correlators (in case of mesons, $\sinh$ in case of baryons) on 64$^4$ lattices. }

\end{table}

\subsubsection{Charm mass tuning}
From ref. \cite{CharmMass} the ratio $m_c/m_s = 11.85$. Since with Wilson fermions, there 
is an additive renormalization, it is not possible to use this ratio directly in setting the
charm mass. However, we know that for $ud$ and $s$ the masses used in the simulation correspond to 
a mass ratio of 1.5 \cite{Durr:2010vn, Durr:2010aw}, from this we get $(m_c-m_s)/(m_s-m_{ud})=35.55$ which gives the estimate
for the charm mass that was used. To check if this is approximately the correct charm mass,
we checked the masses of the different mesons states containing s and c quarks, and they were indeed
in the right ballpark. See Table 2.

\subsubsection{Approximation of the "critical temperature"}
Since the simulation is done at a non-physical pion mass, instead of giving the temperature values in
MeV, it is probably more appropriate to give temperatures in $T_c$ units. Since with dynamical
light quarks there is no phase transition, the transition is a cross-over,
there is a multitude of possible definitions of
$T_c$. We will use the temperature value where the tree-level improved strange quark susceptibility 
equals 0.5. One can obtain an estimate of this temperture from Figure 2 of Ref. \cite{WilsonThermo},
where the continuum extrapolated strange quark susceptibility is plotted with systematic and statistical errors.
The transition temperature is $T/m_{\Omega} = 0.110(2)$.

\section{Results}

\begin{figure}[t!]
\begin{center}
\centering
\includegraphics[angle=270,width=0.8\linewidth]{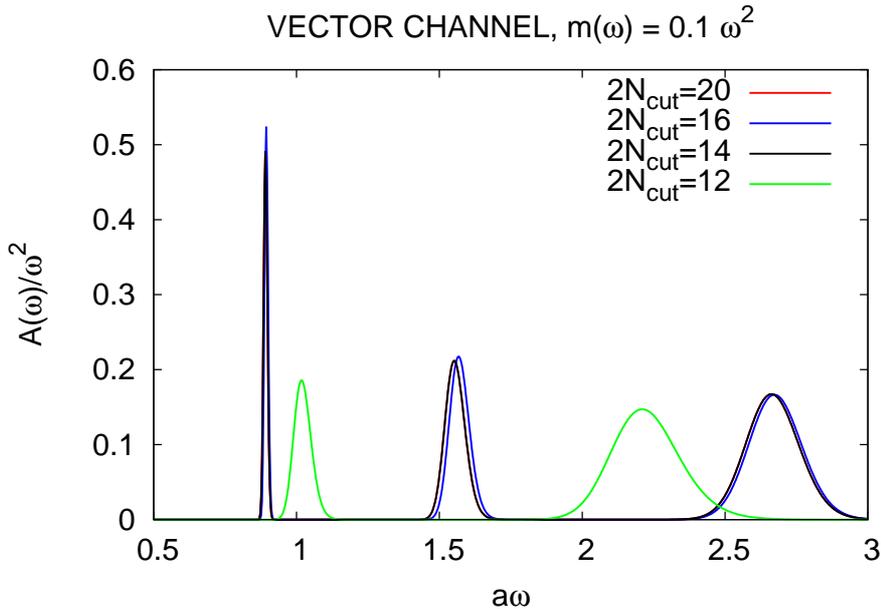}
\renewcommand{\figure}{Fig.}
\caption{The results of dropping points from the zero temperature reconstruction. We dropped points starting from $t = 0$, using only the points 
closest to $t = N_t/2$. $N_{\rm{cut}}$ is the number of points used for reconstruction. For $2N_{\rm{cut}} \leq 12$ one can no longer reconstruct the first peak.}
\label{fig:Tzero}
\end{center}
\end{figure}

\subsection{Zero temperature analysis}
Since the temperature is $T=1/(N_t a)$, as the temperature increases we have less 
and less data points for our reconstruction of the SFs. That means
that the reliability of the method decreases with increasing temperature. So we need an estimate
of the highest temperature, where the MEM results are still likely to be trusted. 
To get such an estimate, we drop points from the lowest temperature correlators
and do a MEM reconstruction with these limited number of points. We say that the
reconstruction is no longer reliable when we can not reconstruct the first peak.
The results are illustrated in Fig. \ref{fig:Tzero}. As we can see, $N_t=12$ is
already not reliable, meaning that the highest temperature we can study with direct MEM
reconstruction is corresponding to $N_t=14$. \footnote{Of course, this is not an absolute criterion.}

\begin{figure}[t!]
\begin{center}
\includegraphics[angle=270,width=0.495\linewidth]{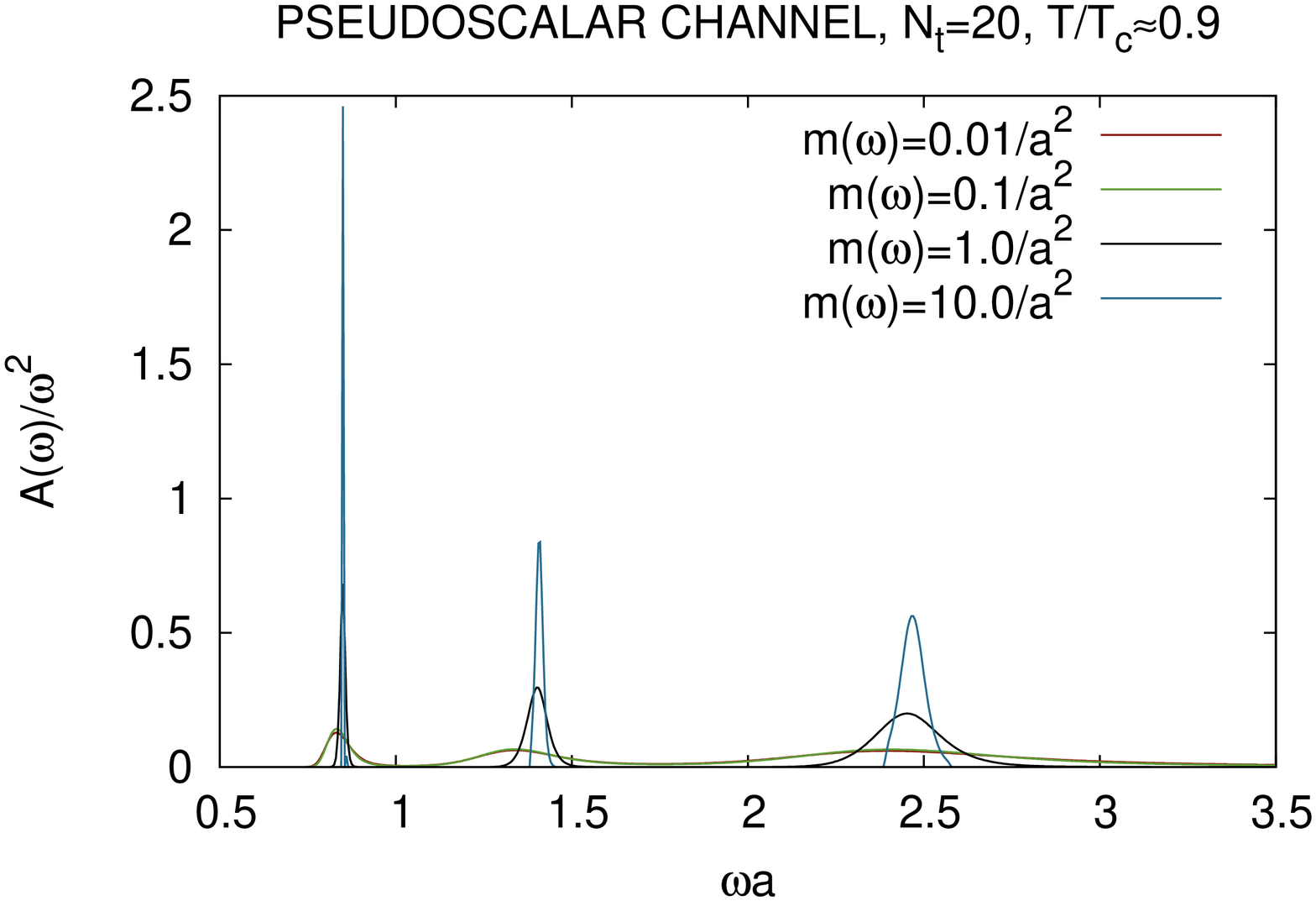} 
\includegraphics[angle=270,width=0.495\linewidth]{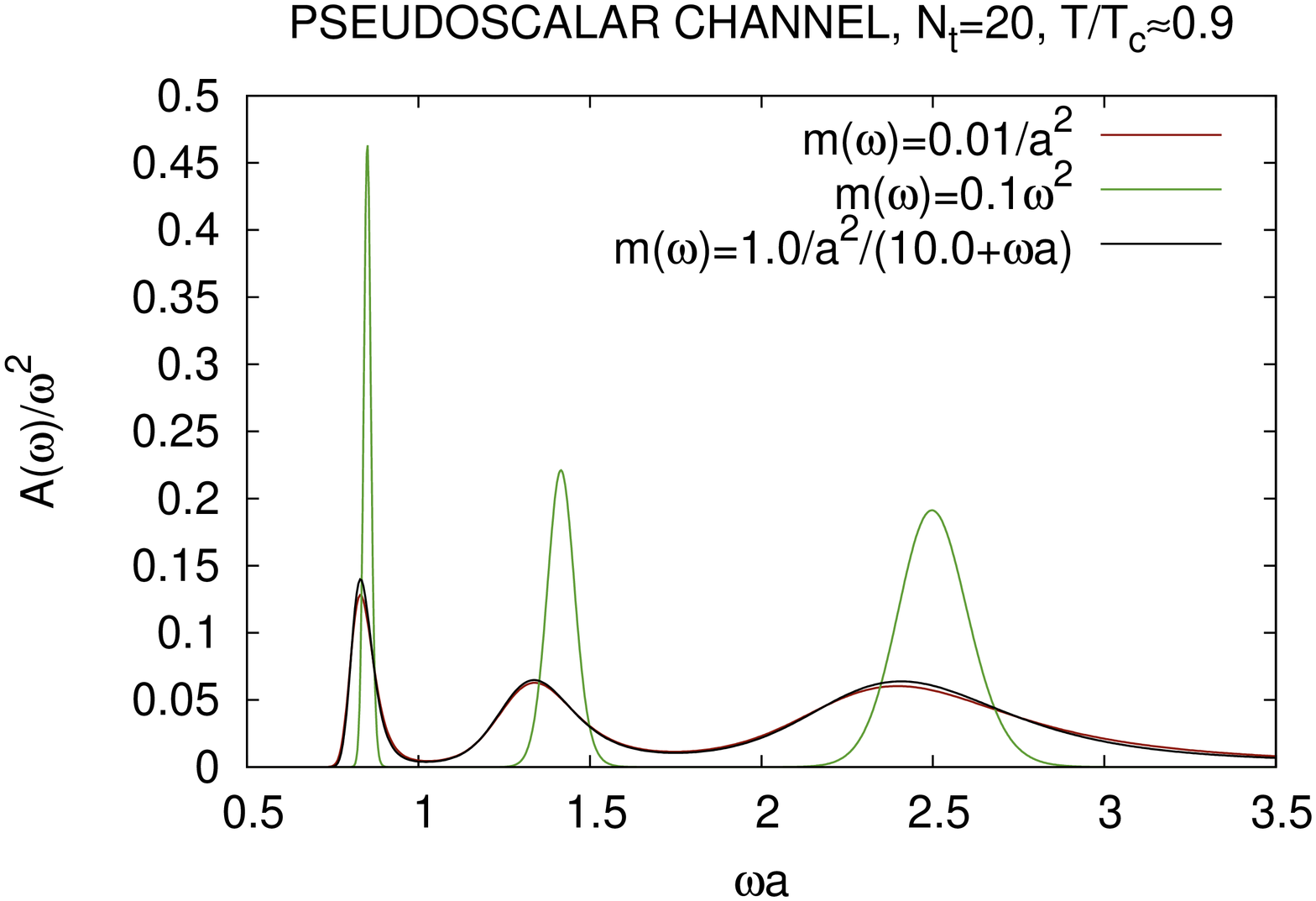} 

\vspace{1cm}

\includegraphics[angle=270,width=0.495\linewidth]{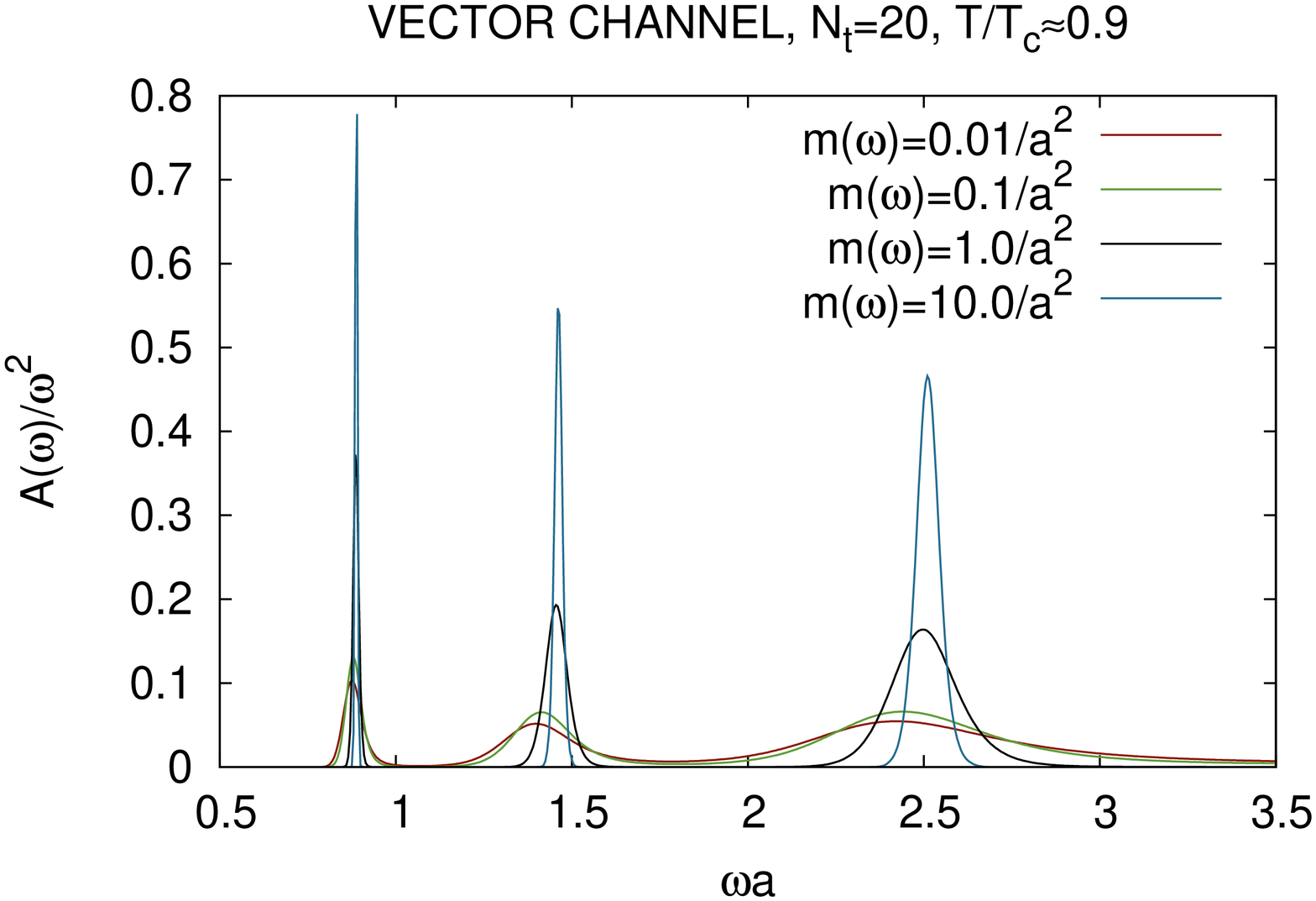} 
\includegraphics[angle=270,width=0.495\linewidth]{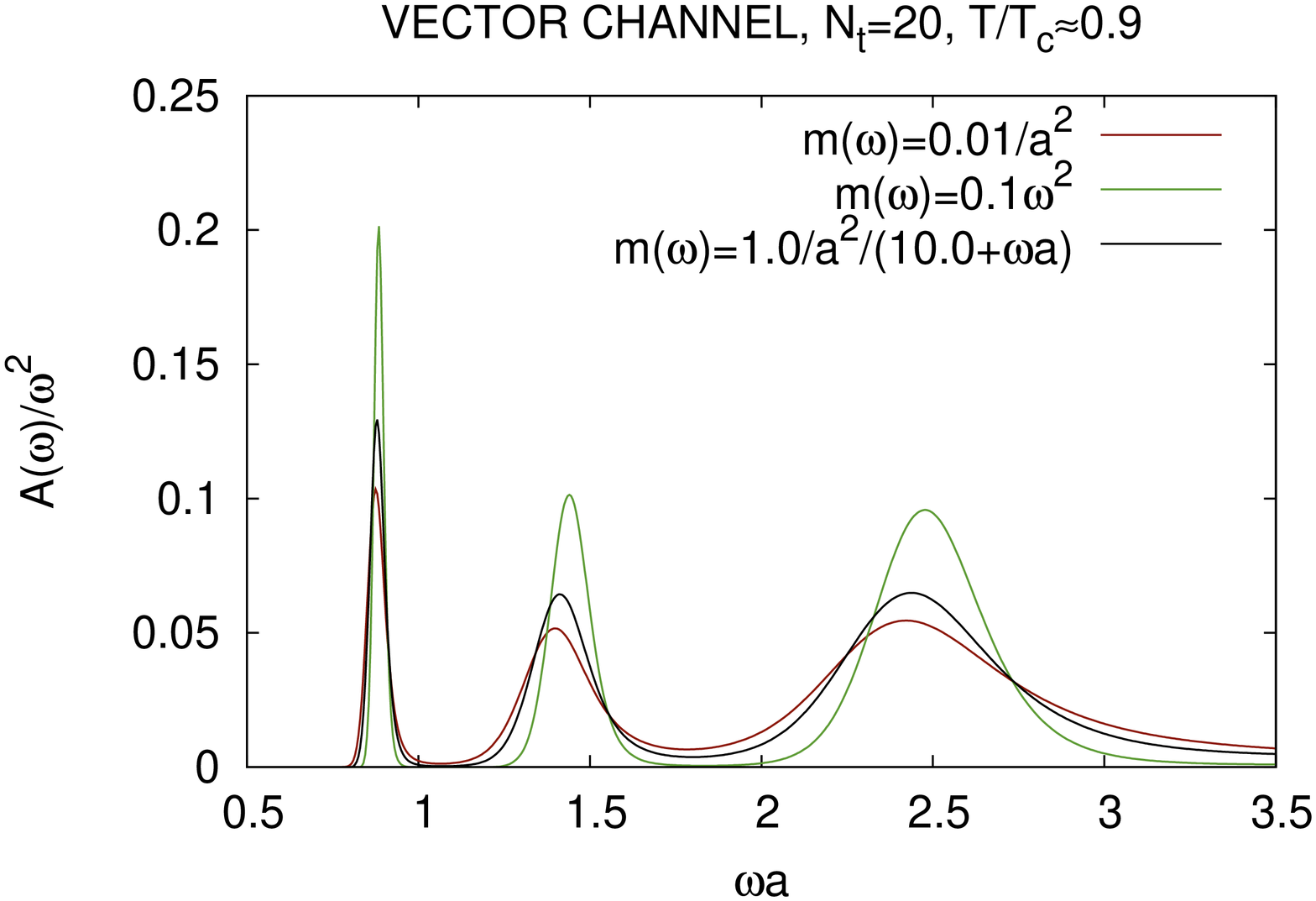}
\end{center}

\renewcommand{\figure}{Fig.}
\caption{ The sensitivity of the reconstruction on the prior function. Note that the widths of the 
peaks have large systematic errors, as can be seen from the reconstruction with different prior functions.}
\label{fig:prior}

\end{figure}

\vspace{2cm}

\begin{figure}[h!]
\begin{center}
\includegraphics[angle=270,width=0.495\linewidth]{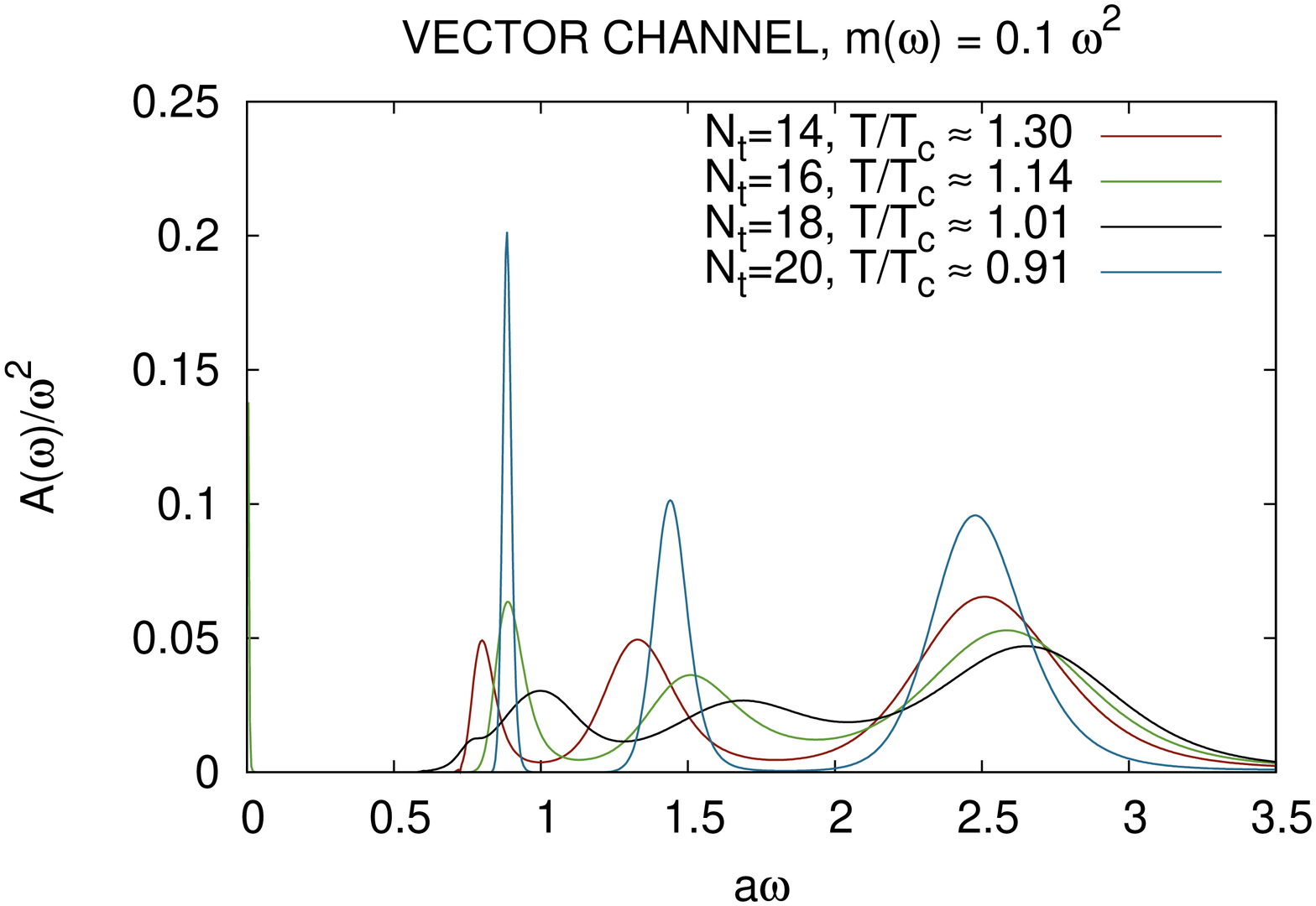} 
\includegraphics[angle=270,width=0.495\linewidth]{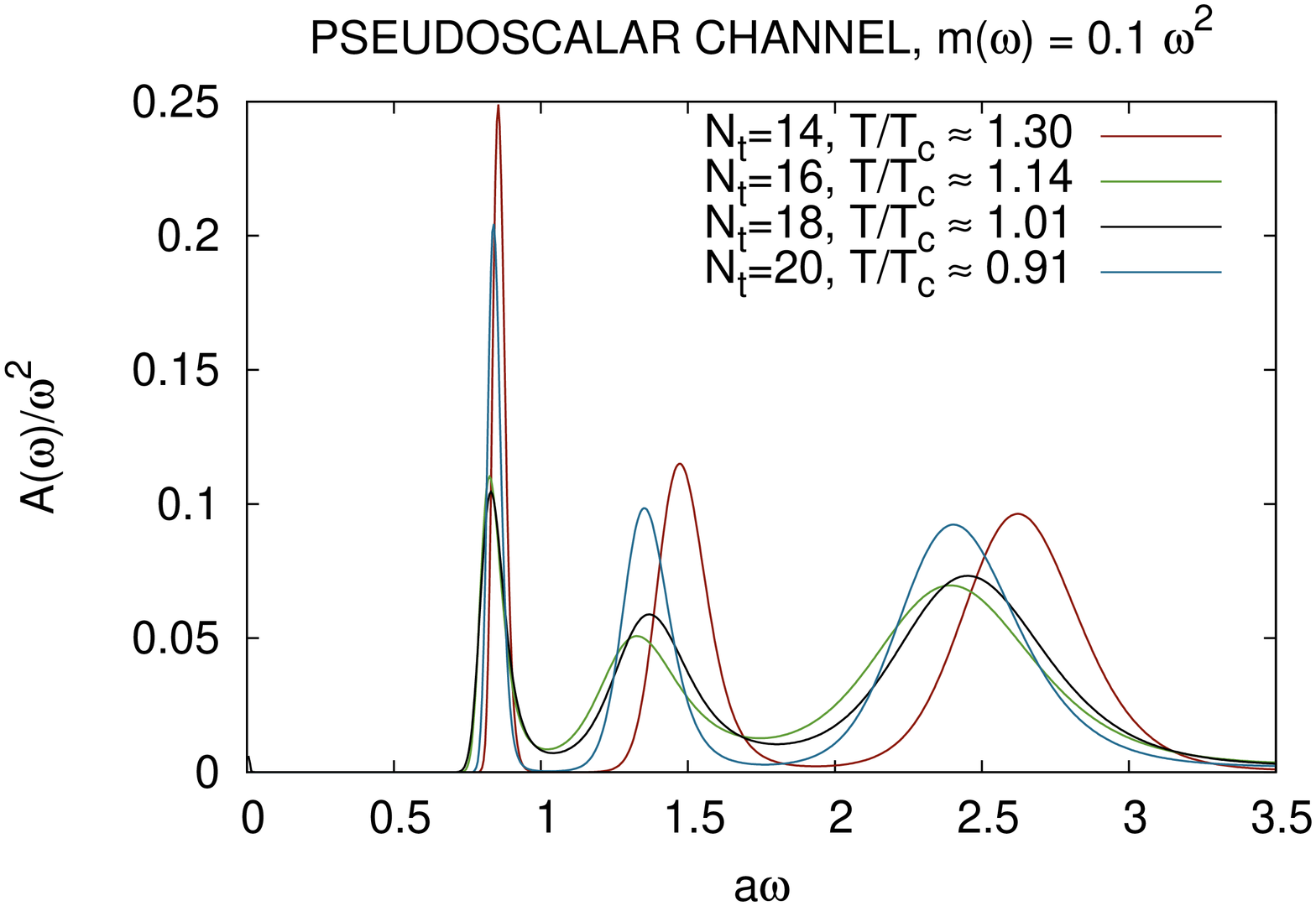} 

\end{center}

\renewcommand{\figure}{Fig.}
\caption{The temperature dependence of the reconstucted spectral functions.}
\label{fig:temp_SF}

\end{figure}

\subsection{MEM reconstructed spectral functions}
\subsubsection{Pseudoscalar channel}

\begin{figure}[t!]
\begin{center}
\centering
\includegraphics[angle=270,width=0.495\linewidth]{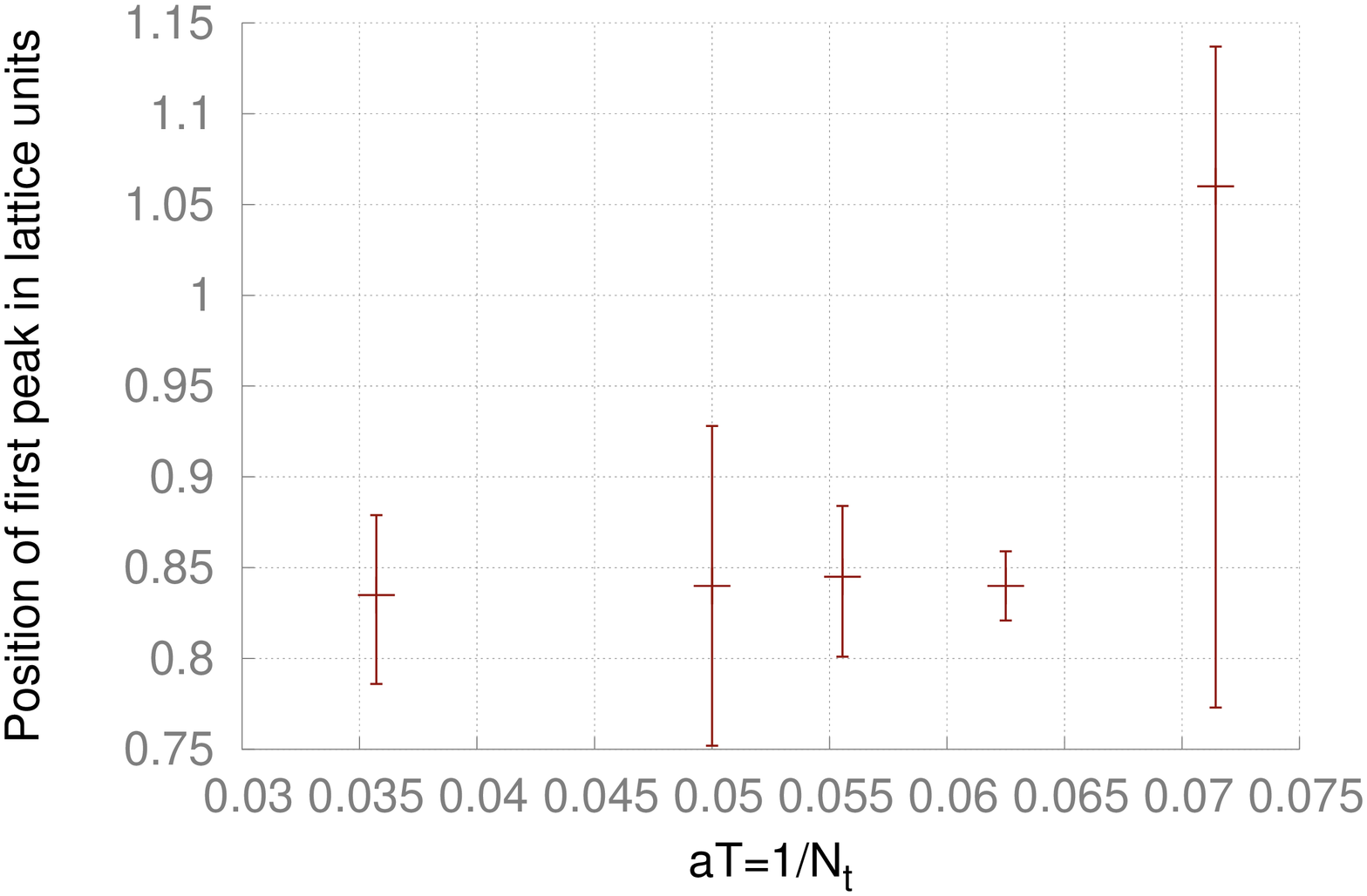}
\includegraphics[angle=270,width=0.495\linewidth]{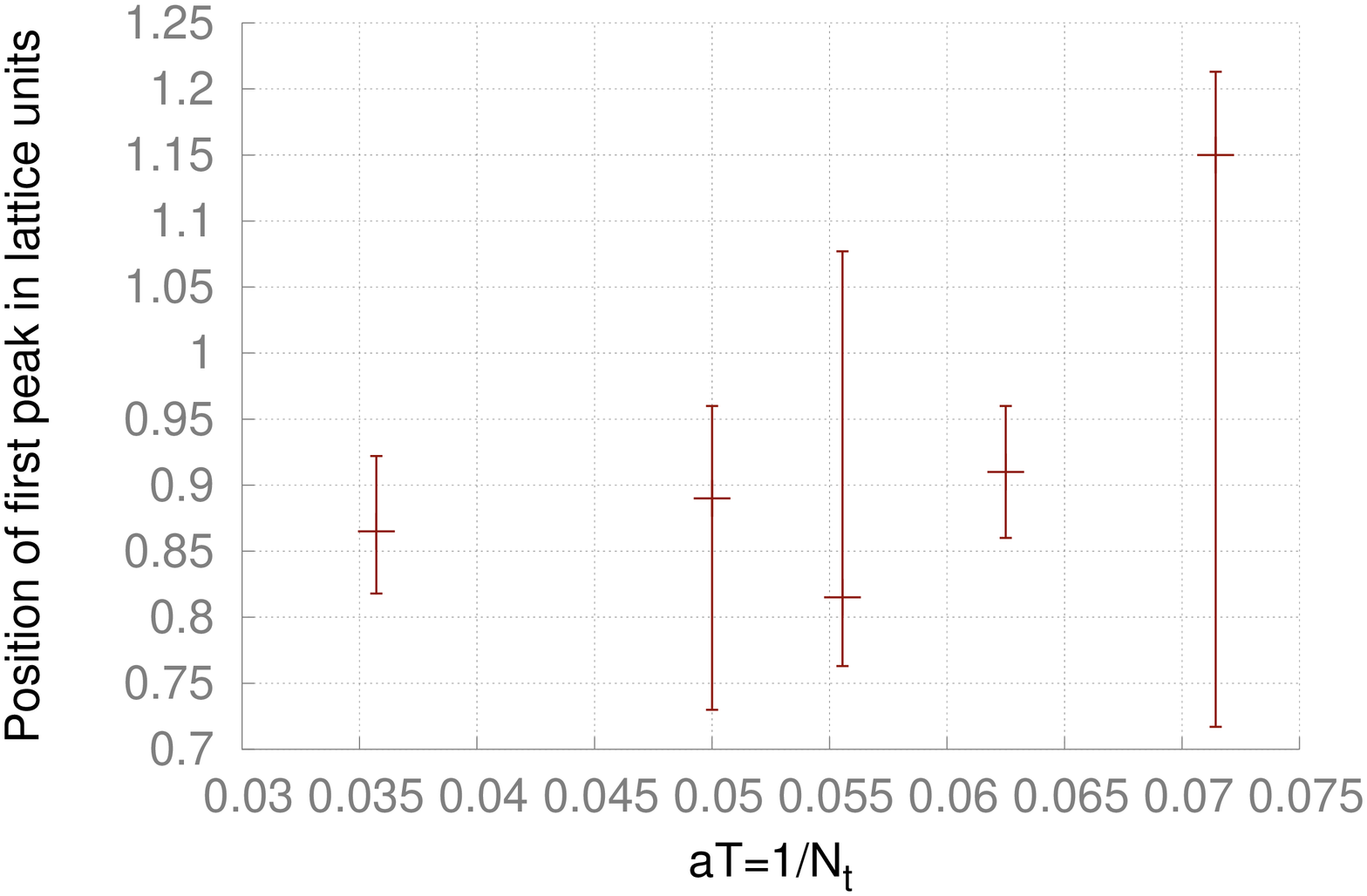}
\renewcommand{\figure}{Fig.}
\caption{The position of the first pseudoscalar(left) and vector(right) peak as a function of temperature. The big error bars at the highest
temperature, corresponding to $N_t=14$ come from the fact that with so few data points, the MEM procedure merges the first two peaks with
some prior functions. Errorbars include both systematic and statistical errors.}
\label{fig:peak_temp}
\end{center}
\end{figure}

Reconstructions of the pseudoscalar (PS) SF with different prior
functions can been seen in Fig. \ref{fig:prior}.
Reconstruction of the PS spectral function with the same prior functions at
different temperatures can be seen in Fig. \ref{fig:temp_SF}. Looking at these
pictures together one can draw the intuitive conclusion that the difference in the 
SFs at various temperatures is smaller than the reconstruction error
coming from the variation of the reconstruction with different prior functions. 
So as far as our analysis can tell, the PS SF is temperature
independent in the given range. This is further confirmed by Fig.
\ref{fig:peak_temp}, which shows a full error
analyis of the peak position.

\begin{figure}[t!]
\begin{center}
\centering
\includegraphics[width=0.495\linewidth]{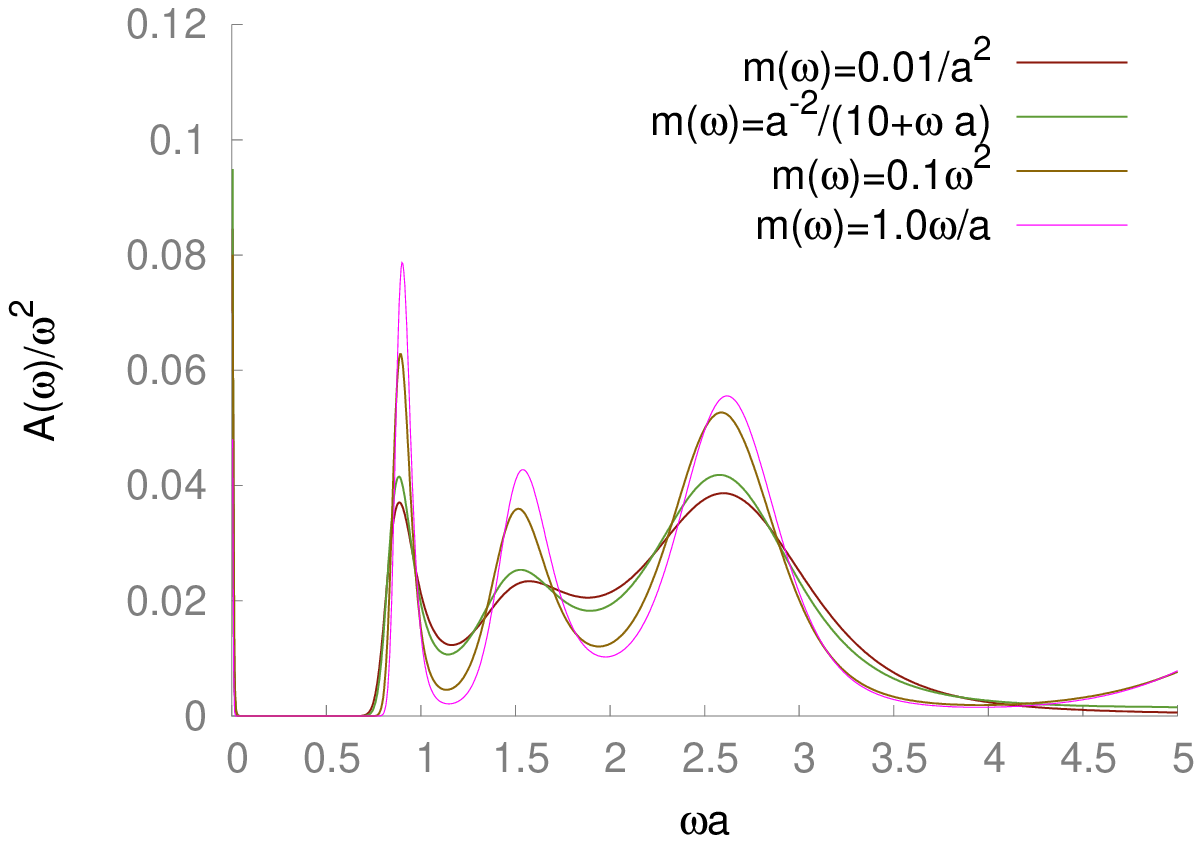}
\includegraphics[width=0.495\linewidth]{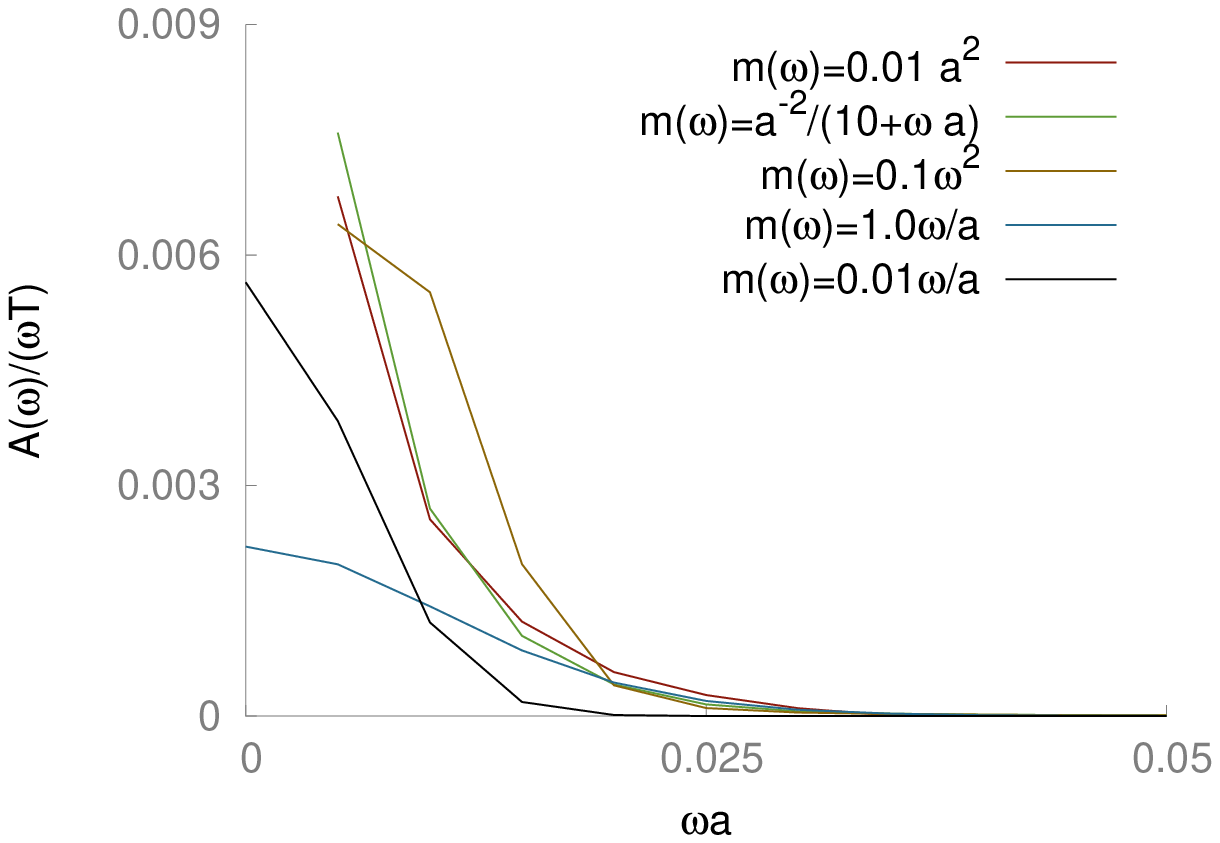}
\renewcommand{\figure}{Fig.}
\caption{The MEM reconstructed vector SF at $N_t=16$ in the vector channel. Notice the 
indication of a transport peak in the MEM analysis. With the prior functions $A(\omega)=m_0 a^{-2}$ or $\frac{a^{-2}}{m_0+\omega a}$
the zero frequency limit of $A(\omega)/\omega$ is always infinity, with $A(\omega)=m_0 \omega^2$ it is always zero.
To actually get a value for the transport coefficients, we need to use the prior function $A(\omega)=m_0 \omega a^{-1}$.
This way we can get an estimate, but as the results show, it has a big systematic uncertainty. As we will stress in
the analysis of the "reconstructed" correlators, the data are not really sensitive to the zero frequency
intercept, only the area of the transport peak.
}
\label{fig:V_nt_16}
\end{center}
\end{figure}

\subsubsection{Vector channel}
The situation is a little bit more complicated in the vector channel. Reconstructions 
of the vector SF with different prior functions can been seen in Fig. \ref{fig:prior}. 
Reconstruction of the vector SF with the same prior functions 
at different temperatures can be seen in Fig. \ref{fig:temp_SF}. In this last plot 
the highest temperature seems to differ from the other temperatures. 
The first peak appears to go down to lower temperatures. Due to some properties of the analysis (i.e.
possible merging of adjacent peaks and problems with the resolution of the transport
peak) using MEM alone one cannot draw any firm conclusions about the
nature of the change in the SF - at least at the current level of statistical errors. At $N_t=16$ the MEM reconstruction
picks up a transport related low frequency peak. However at $N_t=14$ the MEM is
already not reliable in the vector channel. From mock data analysis, we 
observed that MEM can merge two close lying peaks to one peak between the two real peaks.
We suspect that this is what is happening at $N_t=14$. The $N_t=14$ reconstruction also picks up a transport peak 
with some prior functions, but here we don't see three $\omega>0$ peaks together with a transport peak. The first
peak is always merged with the transport peak, or the second peak. This peak merging property makes the error bars on the 
peak position at this temperature so big. 
Fig. \ref{fig:peak_temp} shows a full error analyis of the peak 
position. The actual physical picture will be clarified in the next point of our analysis.

\subsection{The ratio $G/G_{\rm{rec}}$}
An alternative aproach to study spectral functions was suggested in \cite{Jakovac}. The ratio:
\begin{equation}
\frac{G \left( t, T \right) }{G_{\rm{rec}} \left( t, T \right)}  = \frac{ G(t,T) }{ \int A(\omega,T_{\rm{ref}})K(\omega,t,T)\rm{d}\omega }
\end{equation}
has a few advantages:
\begin{itemize}
 \item MEM reconstruction is only needed at $T_{\rm{ref}}$, where we have the most data points, and 
so a more reliable reconstruction. We use $N_t=28$ as reference temperature. \\
 \item We can calculate this ratio even at high temperatures, where the MEM reconstruction is
 already unreliable. \\
 \item If the spectral function is temperature independent, then the 
 trivial temperature dependence of the correlators, coming from the 
 integral kernel will drop out, and the ratio will be $G/G_{\rm{rec}}=1$. \\
\end{itemize}
It is also useful to study the same ratio with mid-point subtracted correlators \cite{Umeda:2007hy}.
\begin{multline}
\frac{G^-}{G^-_{\rm{rec}}} = 
\frac{G \left( t, T \right) - G \left( N_t/2, T \right) }
{G_{\rm{rec}} \left( t, T \right) - G_{\rm{rec}} \left( N_t/2, T \right)}
= \\
\frac{ G \left( t, T \right) - G \left( N_t/2, T \right) }
{ \int A(\omega,T_{\rm{ref}})\left[ K(\omega,t,T)-K(\omega,N_t/2,T) \right] \rm{d}\omega }
\end{multline}
This way, one can drop the zero-mode (constant) contribution to the correlators. These
have to do with transport coefficients, or other low frequency ($\omega \ll T$) features of the spectral
functions. 

\begin{figure}
\begin{center}
\includegraphics[angle=270,width=0.495\linewidth]{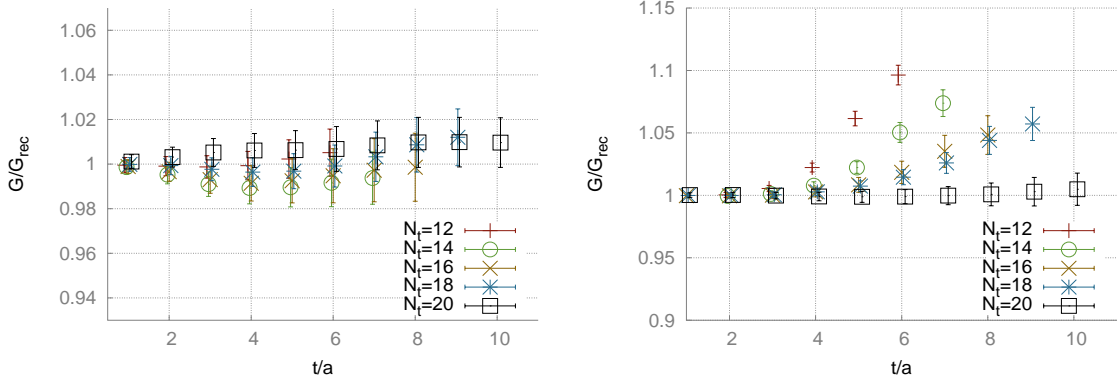} 
\includegraphics[angle=270,width=0.495\linewidth]{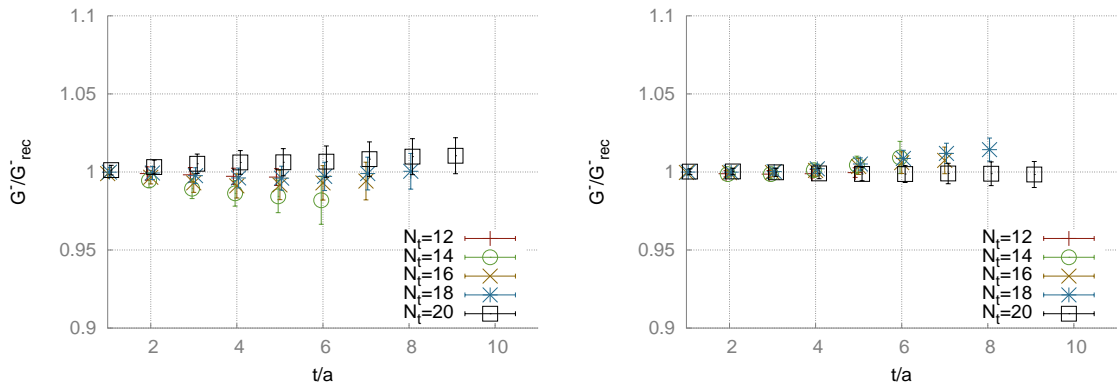} 
\caption{The ratio $G/G_{\rm{rec}}$ in the pseudoscalar(left) and vector(right) channels. Errorbars include both systematic and statistical errors.}
\label{fig:GGrec}
\end{center}
\end{figure}

\begin{figure}
\begin{center}
\includegraphics[angle=270,width=0.495\linewidth]{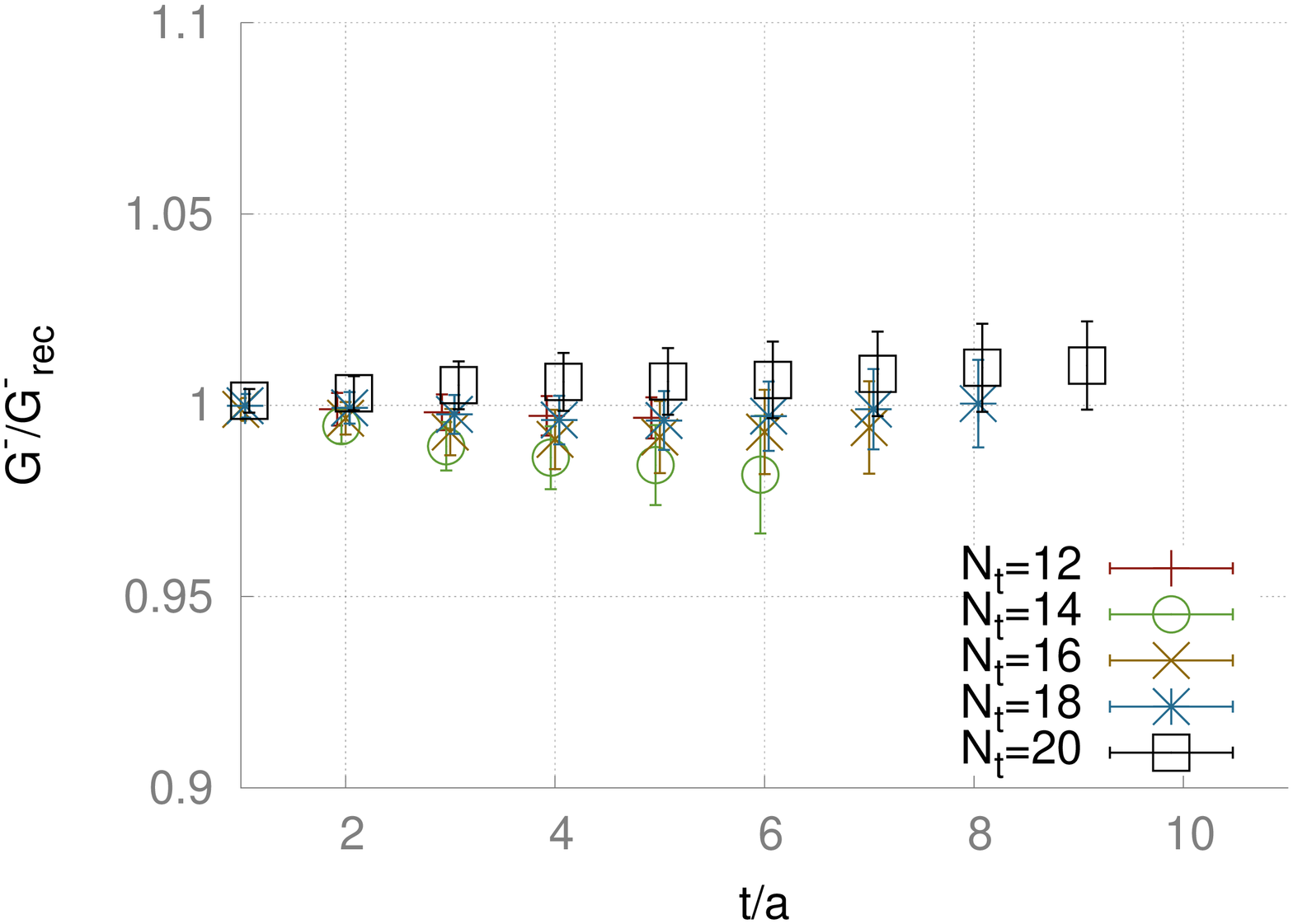} 
\includegraphics[angle=270,width=0.495\linewidth]{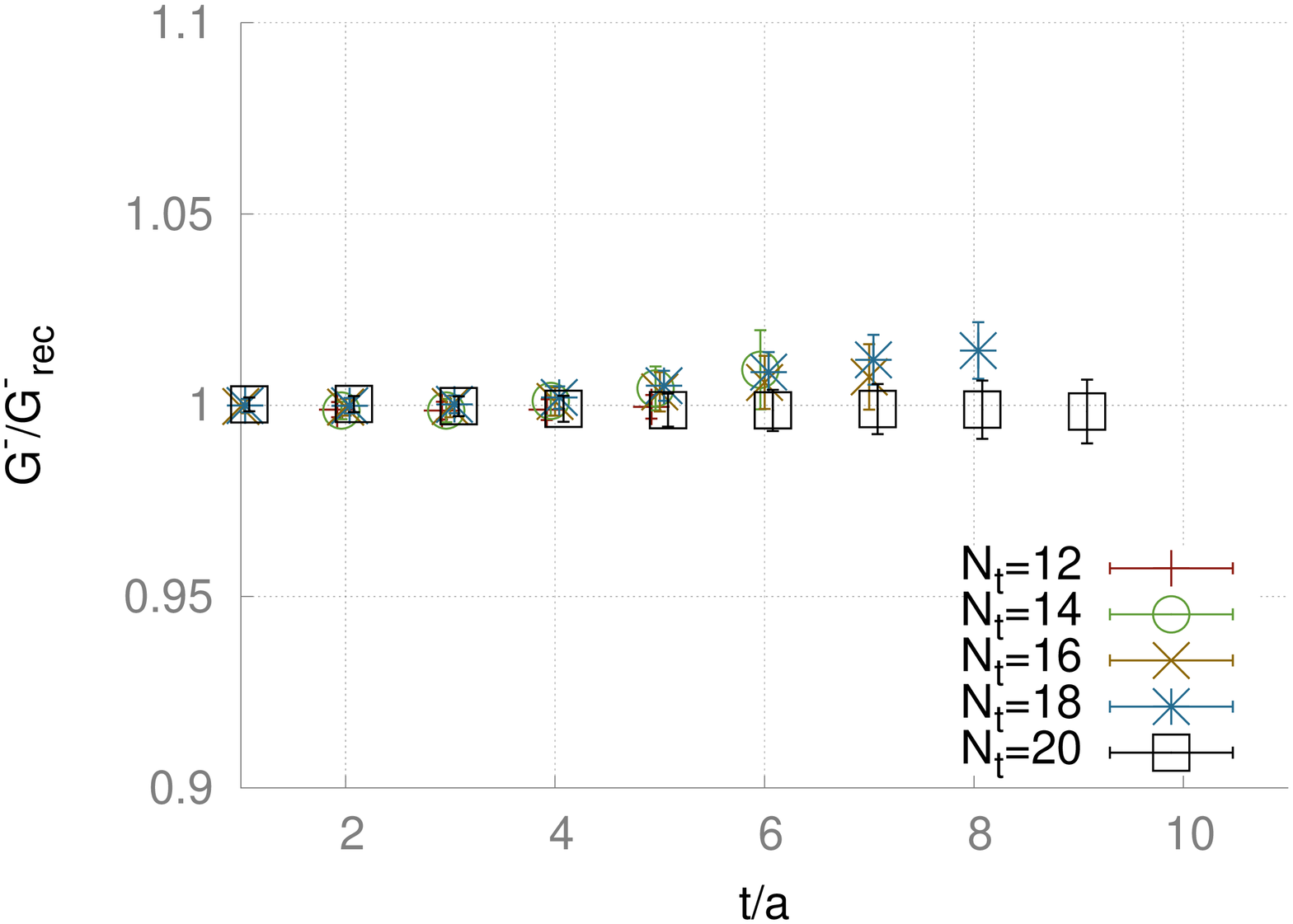} 
\caption{The ratio $G^{-}/G^{-}_{\rm{rec}}$ in the pseudoscalar(left) and vector(right) channels. Errorbars include both systematic and statistical errors.}
\label{fig:GGrec_sub}
\end{center}
\end{figure}

\begin{figure}[t!]
\begin{center}
\centering
\includegraphics[angle=270,width=0.8\linewidth]{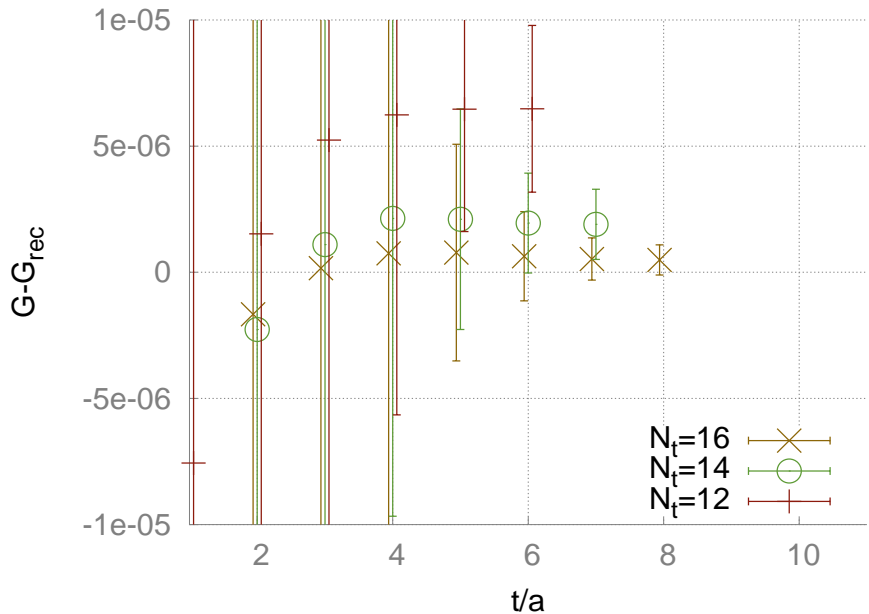}
\renewcommand{\figure}{Fig.}
\caption{The difference $G-G_{\rm{rec}}$ in the vector channel at the 3 highest temperatures. If we 
assume the ansatz $A(\omega,T)=f(T)\omega \delta(\omega-0^+)+A(\omega,T_0)$ the difference is equal to
$f(T) T / 2$. Errorbars include both systematic and statistical errors.}
\label{fig:G_min_Grec}
\end{center}
\end{figure}

If the ratio of $G/G_{\rm{rec}}$ is
different from one, but the ratio with the middle-point substracted correlators is not, that
means that the temperature dependence of the SFs should be well described by just a zero-mode
contribution $f(T) \cdot \omega \delta(\omega-0^+)$. The results of such an analysis can be seen
in Figs. \ref{fig:GGrec} and \ref{fig:GGrec_sub}. As one can see, the results in the pseudoscalar
channel are consistent with a temperature independent SF, while the results in the vector channel
show a temperture dependent zero mode/low frequency contribution in the SF. We can also try to 
extract the zero mode contribution itself by considering the difference $G-G_{\rm{rec}}$. This is only
plotted in the vector channel, in Fig. \ref{fig:G_min_Grec} (in the pseudovector channel it is always 
consistent with zero). The difference has big errors, but on the two highest temperatures it is 
non zero within 1$\sigma$. At every temperature it is consistent with a time separation
independent constant. With the ansatz $A(\omega,T)=f(T)\omega \delta(\omega-0^+)+A(\omega,T_0)$
we get $f(T)T \approx (3 \pm 1.5)\cdot 10^{-5}$ at $1.5T_c$ in lattice units \footnote{Since we are not
using the conserved current on the lattice, but a local current, this will have a finite, lattice spacing dependent
renormalization constant of $\mathcal{O}(1)$. We neglect this fact, since we don't do a continuum limit, and the
renormalization is temperature independent. }. This ansatz, taken strictly, would imply a diverging diffusion
constant. However, the data do not restrict the shape of the transport peak, they are only
sensitive to the area. By using this ansatz, we do not mean to say that the diffusion constant diverges,
we simply extract that area of the transport peak. To get a diffusion constant additional information is needed.
(The width or the height of the peak, which is too narrow to resolve at this point.) 
The survival of $J/\Psi$ up to such high temperatures is consistent with previous results in 
quenched and 2 flavour QCD (see eg. \cite{Ohno11, Ding, Kelly:2013cpa}). \\

\section{Summary and outlook}

We have performed a lattice study of charmonium spectral functions
with 2+1 dynamical Wilson quarks. The MEM reconstruction of the spectral functions is
hampered by the limited number of data points at higher temperatures, so the highest
temperature we used for MEM reconstruction was approximately $1.3T_c$.
The PS spectral functions did not show any noticable temperature variation. The V spectral
functions showed a temperature variation, which, according to the analysis of the ratio 
$G/G_{\rm{rec}}$ is consistent with a temperature dependent zero mode, and a temperature
independent non-zero part in the SFs. In conclusion, we can say that we observed no 
melting of the $\eta_c$ and $J/\Psi$ mesons up to temperature of $1.5T_c$, and we observe
no variations in the spectral functions of $\eta_c$ whatsoever. The variations in
vector SF can be well described by the ansatz $A(\omega,T)=f(T)\omega \delta(\omega-0^+)+A(\omega,T_0)$
with $f(T)T \approx (3 \pm 1.5)\cdot 10^{-5}$ at $1.5T_c$ in lattice units, giving the area of the transport
peak.\\

Mock data analysis shows that the errors of the data points 
are more important than the number of data points itself (see e.g. \cite{Asakawa}), so 
anisotropic lattices are not expected to substantially increase the accuracy. However
because of the increased number of data point at higher temperatures, the reconstruction
is likely to be reliable at higher temperatures than with isotropic lattices.
Anisotropy tuning with dynamical fermions is a difficult task, but it can somewhat 
be made easier by the method described in \cite{wflow}. This will be the main direction 
for future studies.

\section*{Acknowledgment}

We thank P. Petreczky and A. Jakov\'ac for useful discussions.\\

Computations were carried out on GPU clusters \cite{Egri:2006zm} at the
Universities of Wuppertal and Budapest as well as on supercomputers in
Forschungszentrum Juelich. \\

This work was supported by the EU Framework Programme 7 grant (FP7/2007-2013)/ERC No 208740, by the Deutsche Forschungsgemeinschaft
grants FO 502/2, SFB-TR 55 and by Hungarian Scientific Research Fund grant OTKA-NF-104034.. \\

\section*{Appendix A - MEM details}

\subsection*{Numerical implementation}

The statistics currently feasible for dynamical Wilson-fermion calculations are very small compared to 
quenched simulations, meaning additional care has to be taken in the MEM analysis. We mention here the
difficulties, that were not stressed earlier in the literature, summarizing our experience. \\

First of all, for realistic data double precision is not enough to make a reliable reconstruction
of the SFs. Even in the case of the algorithm of Ref. \cite{Jakovac}, which does not involve a 
Singular Value Decomposition, the Hessian matrix of the function $U$ has a big condition number, 
and arbitrary precision arithmetics is needed to find the correct minimum. Our implementation uses 
the Levenberg-Marquardt (or optionally the LBFGS) algorithm in arbitrary precision, implemeted
by help of the GNU Multiple Precision Library. Using too small precision will generally lead 
to not finding the correct minimum. In practice, we have witnessed that using smaller precision leads 
to broader peaks. \\

Also, even with the high precision one has to be very careful about the stopping criteria for the iteration.
The general behaviour is that after a rapid decrease of the function U, further iterations
hardly improve it, i.e. it looks like hitting a plateau. Then after quite a few iterations, it starts improving fast 
once again, creating a step like
pattern.  With such a behaviour, one has to choose a very strict stopping criteria. \\

\begin{figure}[t!]

\begin{center}

\includegraphics[angle=270,width=0.495\linewidth]{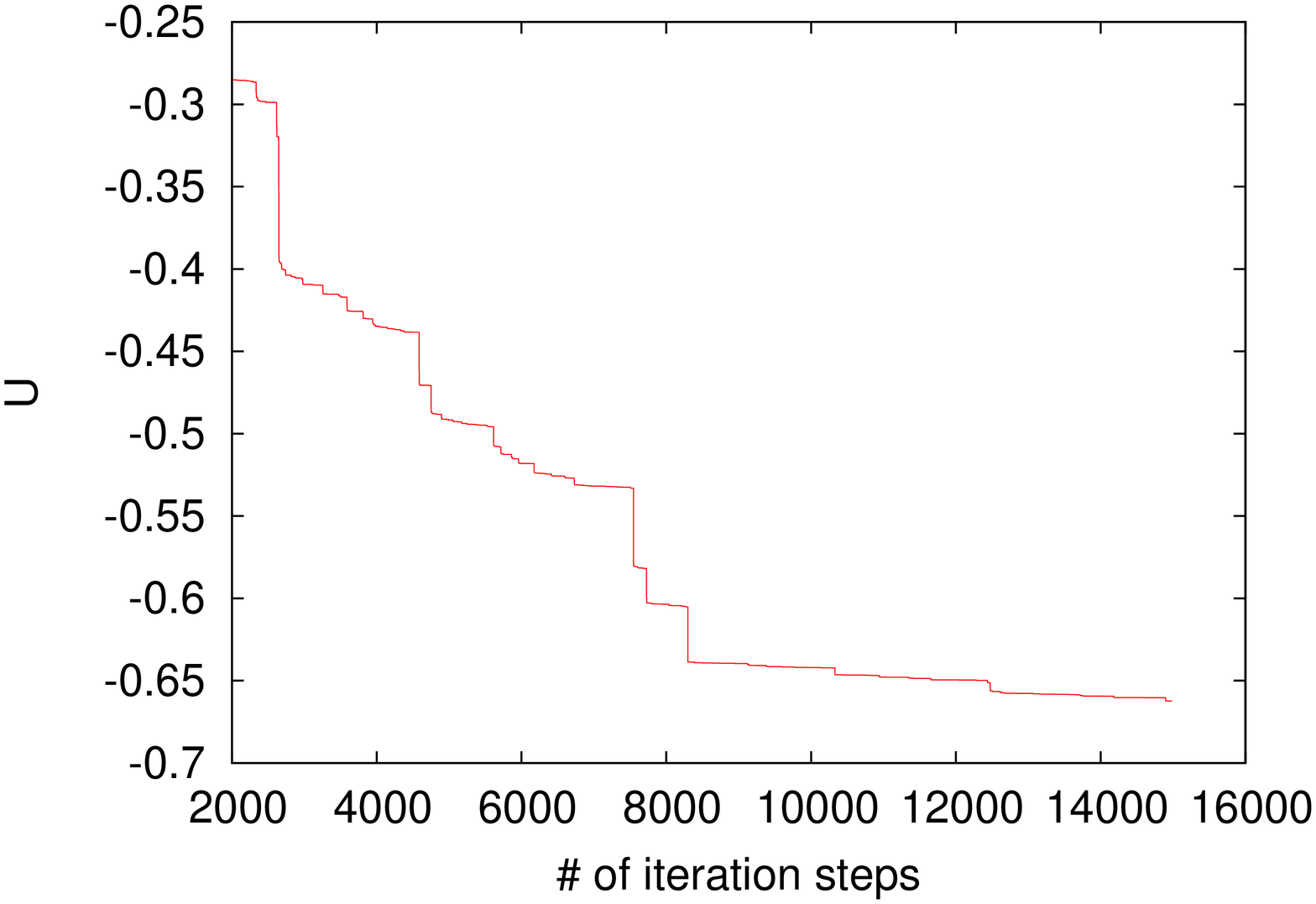}
\includegraphics[angle=270,width=0.495\linewidth]{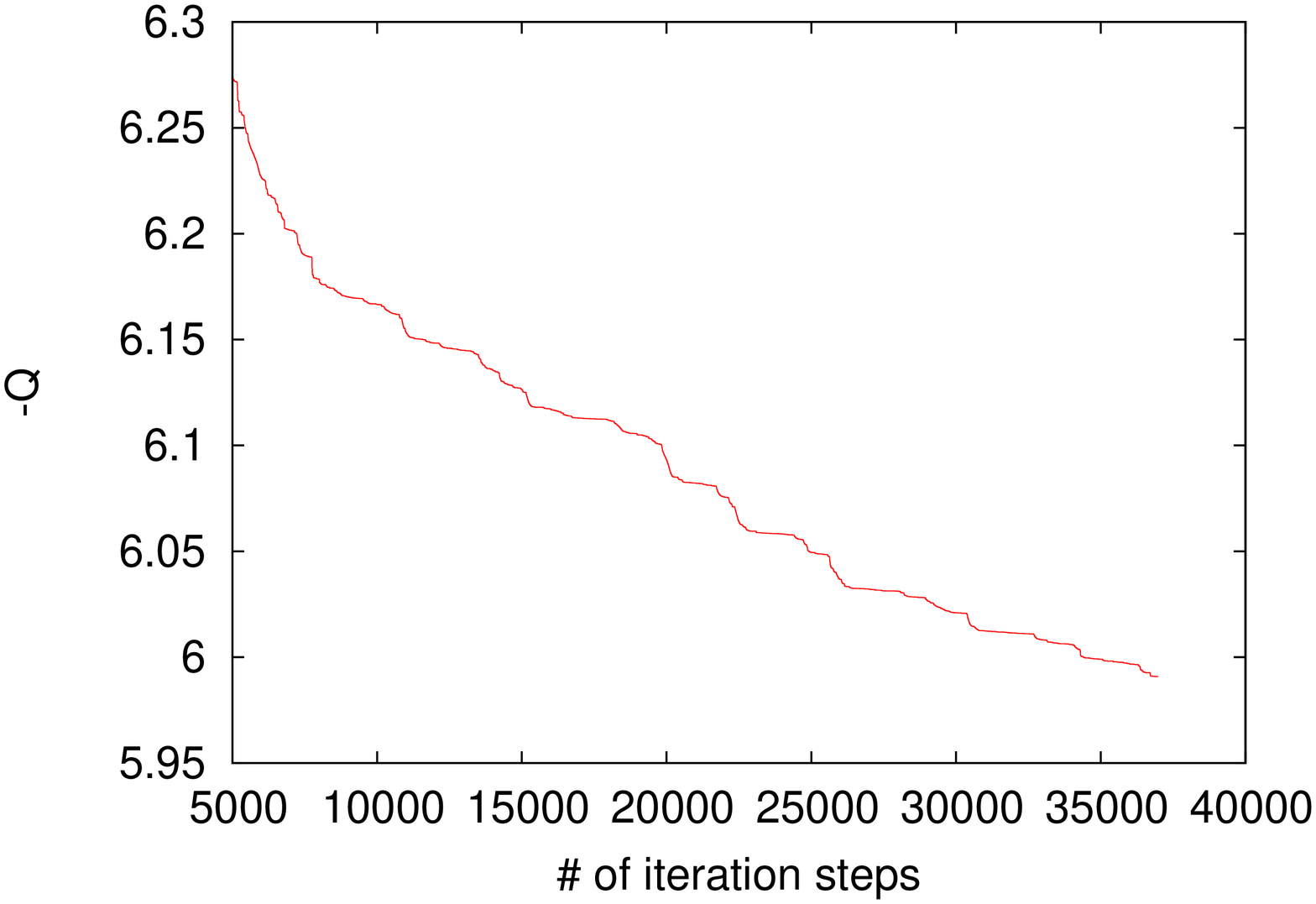}

\end{center}
\caption{The behaviour of the objective functions U and -Q (eqn. (\ref{eq:Q}) and (\ref{eq:U})) as a function of
the iteration steps in the $N_{\rm{data}}$(left) and $N_{\omega}$(right) dimensional spaces.}
\label{fig:algo}
\end{figure}

If we work in the original $N_\omega$ dimensional space (here, we can use the LBFGS algorithm), 
and maximize Q instead, this situation is slighly better. In that space, the behaviour is not step-like, 
but an iteration takes much more time, and whatever method one chooses, it takes lots of iterations 
to find the minimum, making the analysis computationally costly. This behaviour is illustrated 
on Fig. \ref{fig:algo}.\\

In the end, we decided to follow the following procedure: minimize $U$ in the $N_{\rm{data}}$ dimensional space,
then after the iteration stops switch to the $N_{\omega}$ dimensional space and do some iterations to
check if we found the true minimum, or not. \\

\subsection*{Error analysis}
We don't carry out an error analysis of the full spectral function, since with the current 
statistics that would give huge errors. Instead, we only give errors to some physically intersting 
quantities related to the spectral function, that are more stable. \\

The statistical error analysis is done with the usual jackknife method. For the sake of reducing computational cost,
the statistical error estimate was only carried out on a given set of the parameters of the reconstruction algorithm.The
systematic error analysis is carried out by varying the parameters of the reconstruction algorithm. Namely:
\begin{itemize}
  \item the discretization of the frequency variable $\Delta\omega$ \\
  \item the upper cut-off on the integral (\ref{eq:IntegralTransform}), $\omega_{\rm{max}}=N_\omega \Delta \omega$ \\
  \item the shape and normalization of the prior function 
\end{itemize}
A numerical check on the lattice data and mock data analysis show that as long as $\Delta \omega$ is sufficiently small to resolve
the peaks and 
$N_\omega$ is such that $\omega_{\rm{max}}$
is sufficiently big, which means somewhere around 5 inverse lattice spacings, the results are not affected by the choice of
the first two parameters. In this analysis we used $a \omega_{\rm{max}} = 5$ and  $a \Delta \omega = 0.005$. 
The effect of the prior function however, can not be neglected. For the systematical error analysis we
used 3 different shapes:
\begin{itemize}
  \item $m_0 / a^2$, motivated by the philoshopy of "we know nothing".\\
  \item $m_0 \omega^2$, motivated by continuum perturbation theory. We must stress however, that there is no reason to think that
  such an asymptotic behaviour can actually be seen on the lattice. In fact, analytical calculations of the SFs 
  with free Wilson quarks show a different behaviour. (See \cite{FreeQuarkSPF1} and \cite{FreeQuarkSPF2} ) \\
  \item $a^{-3}/(m_0/a+\omega)$, a theoretically unmotivated form. By using such a prior, we try to restrict the reconstructed shapes to the ones 
  that are actually dictated by the lattice data.
  \item $m_0 \omega / a$, motivated by the Kubo formula (\ref{equ:Kubo}). This is the only prior function in the study that allows for a finite, nonzero
  diffusion constant, the others would imply either 0, or infinity. We have only used this prior in the vector channel, where the analysis with the other
  prior functions suggested a transport peak.
\end{itemize}
In all cases, $a$ is the lattice spacing, and $m_0$ was varied between $10.0$,$1.0$,$0.1$ and $0.01$ to  
estimate the systematic errors. The systematic error was taken to be between the 17\% and 83\% percentiles of the 
sorted reconstructed parameters. The final errorbars on the plots include both systematic and statistical errors.\\

We mention, that it has been suggested (see e.g. \cite{Ding}), that the free Wilson fermion SFs should be used as the prior information,
we see no reason to do this however, since the free results have $\mathcal{O}(1)$ corrections in lattice perturbation
theory, and treat the prior as a source of uncertainty instead. \\


\end{document}